\documentclass[runningheads]{llncs}
\usepackage{amsmath, amssymb, amsfonts}

\usepackage{booktabs} 
\usepackage{array}    
\usepackage{multirow} 

\usepackage[table]{xcolor} 
\usepackage{hyperref}      
\hypersetup{
    colorlinks=true,       
    linkcolor=blue,        
    citecolor=blue,        
    urlcolor=blue          
}

\usepackage[T1]{fontenc}   
\usepackage[utf8]{inputenc} 
\usepackage{tcolorbox}
\usepackage{caption}      
\usepackage{subcaption}   
\usepackage{tabularx}    
\usepackage{colortbl}    
\usepackage{siunitx}
\usepackage{adjustbox}
\usepackage{tabularx}
\usepackage{xcolor}
\usepackage{colortbl}    
\newcolumntype{C}{>{\centering\arraybackslash}X}
\usepackage{threeparttable}  
\definecolor{lightblue}{HTML}{D2E9FF}
\definecolor{lightred}{HTML}{FFCCCC}
\definecolor{lightgray}{HTML}{F2F2F2}
\definecolor{headergray}{RGB}{224,224,224}
\definecolor{highlightblue}{RGB}{173, 216, 230}
\definecolor{highlightgreen}{RGB}{144,238,144}

\definecolor{lightblue}{HTML}{D2E9FF} 
\definecolor{lightred}{HTML}{FFCCCC}   
\definecolor{lightgray}{HTML}{F2F2F2}  

\definecolor{lightgray}{gray}{0.9}
\definecolor{lightblue}{RGB}{173, 216, 230}
\definecolor{lightred}{RGB}{255, 182, 193}
\definecolor{headergray}{gray}{0.9}
\definecolor{rowgray}{gray}{0.95}

\definecolor{headerblue}{RGB}{91,155,213}
\definecolor{rowgray}{RGB}{240,240,240}

\usepackage{graphicx}

\begin{document}
\title{Poison-RAG: Adversarial Data Poisoning Attacks on Retrieval-Augmented Generation in Recommender Systems}
\titlerunning{Poison-RAG: Adversarial Data Poisoning Attacks on RAG}
%


\author{Fatemeh Nazary\orcidID{0000-0002-6683-9453}\thanks{Corresponding author: Fatemeh Nazary, email: fatemeh.nazary@poliba.it} \and
Yashar Deldjoo\orcidID{0000-0002-6767-358X} \and
Tommaso di Noia\orcidID{0000-0002-0939-5462}}

\authorrunning{F. Nazary et al.}
%
\institute{Polytechnic University of Bari, Bari, Italy \\
\email{firstname.lastname@poliba.it}}
%
%

\maketitle

\begin{abstract}
This study presents \texttt{Poison-RAG}, a framework for adversarial data poisoning attacks targeting retrieval-augmented generation (RAG)-based recommender systems. \texttt{Poison-RAG} manipulates item metadata, such as tags and descriptions, to influence recommendation outcomes. Using item metadata generated through a large language model (LLM) and embeddings derived via the OpenAI API, we explore the impact of adversarial poisoning attacks on provider-side, where attacks are designed to promote long-tail items and demote popular ones. Two attack strategies are proposed: \textbf{local modifications}, which personalize tags for each item using BERT embeddings, and \textbf{global modifications}, applying uniform tags across the dataset. Experiments conducted on the MovieLens dataset in a black-box setting reveal that local strategies improve manipulation effectiveness by up to 50\%, while global strategies risk boosting already popular items. The results indicate that popular items are more susceptible to attacks, whereas long-tail items are harder to manipulate. Approximately 70\% of items lack tags, presenting a cold-start challenge; data augmentation and synthesis are proposed as potential defense mechanisms to enhance RAG-based systems' resilience. The findings emphasize the need for robust metadata management to safeguard recommendation frameworks. Code and data are available at \href{https://github.com/atenanaz/Poison-RAG}{https://github.com/atenanaz/Poison-RAG}.


\end{abstract}

\section{Introduction}

Retrieval-augmented generation (RAG)~\cite{deldjoo2024recommendation,deldjoo2024review} has emerged as a powerful paradigm in information retrieval and recommendation systems, combining the precision of traditional information retrieval (IR) with the generative capabilities of large language models (LLMs). Recent studies indicate that approximately 60\% of applications utilizing LLMs now integrate RAG frameworks to address issues such as hallucinations and factual inaccuracies inherent in standalone LLMs~\cite{towardsai2024taxonomy}.

\begin{figure}[!t]
    \centering
    \includegraphics[width=1.02\linewidth]{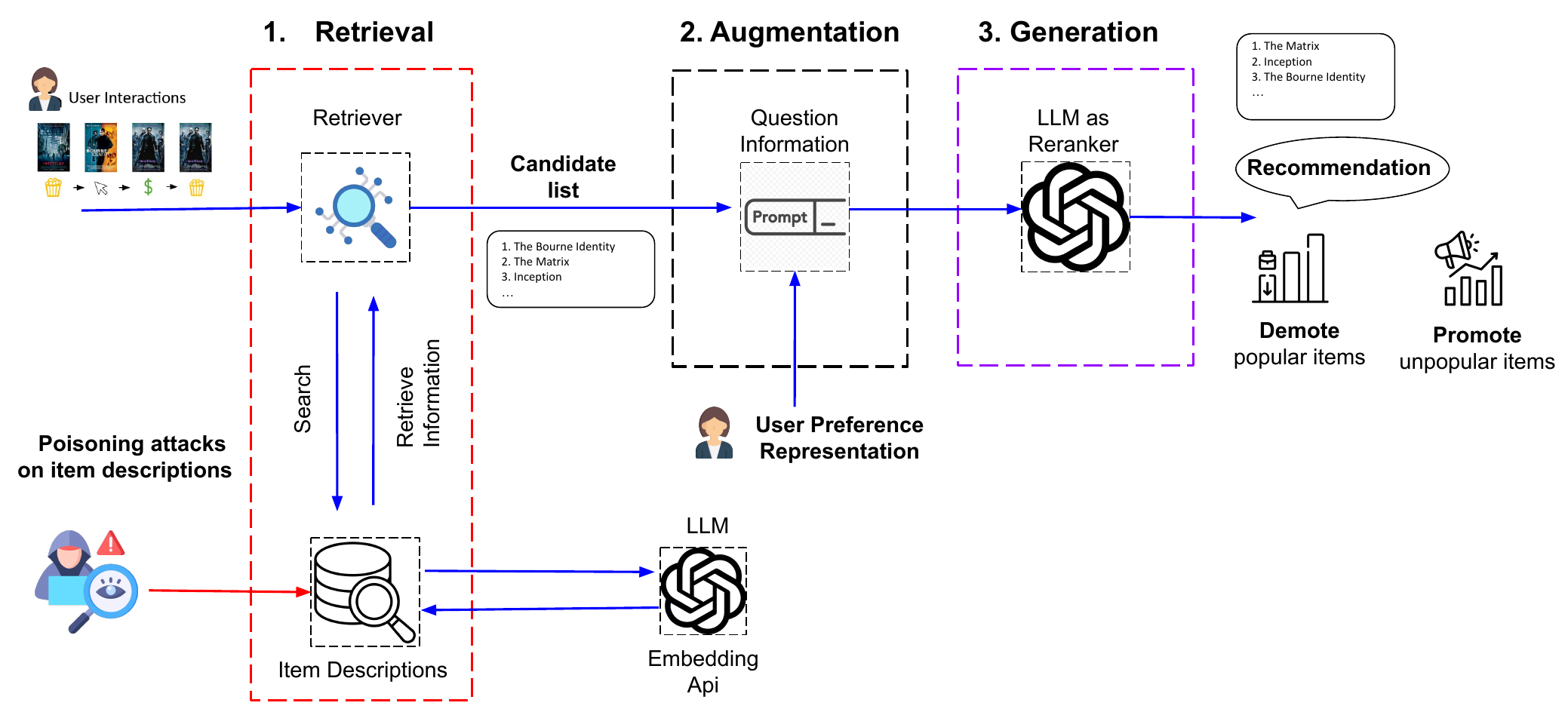}
    \caption{Diagram of the RAG-based recommender system pipeline showing Retrieval, Augmentation, and Generation stages. Adversarial attacks poison item descriptions during Retrieval, aiming to promote unpopular items and demote popular ones in the final recommendations.}
    \label{fig:rag}
\end{figure}

As depicted in Fig.~\ref{fig:rag}, RAG systems typically comprise three stages: \textit{retrieval}, \textit{augmentation}, and \textit{generation}. In the \textit{retrieval} stage, the system accesses an external knowledge base—such as item metadata or user interaction histories—to identify pertinent information. The \textit{augmentation} stage enriches the user query with the retrieved data, providing enhanced context. Finally, during the \textit{generation} stage, the LLM produces responses that are contextually relevant and grounded in factual data, thereby mitigating the risk of generating hallucinations or inaccuracies.

Implementing RAG systems in real-world applications requires careful consideration of each stage to ensure optimal performance. In recommender systems (RS), various RAG applications have been developed based on differing retrieval and augmentation strategies~\cite{di2023retrieval,fan2024survey,lin2023can,deldjoo2024review}. For instance, a recommender system may retrieve candidate items using user-item interaction history in a \textit{collaborative} manner or by leveraging item metadata (e.g., tags, genres, or reviews) to build \textit{textual user profiles}. Each approach generates an initial ranked list, which is then refined during the augmentation and generation steps.

This study focuses on \textit{text-based retrieval} models within RAG systems for two primary reasons: \textit{(i)} they create rich, semantic representations of user interests and item attributes, and \textit{(ii)} they integrate seamlessly with LLMs. Text representations capture nuanced relationships and contextual details that traditional collaborative methods may overlook. Moreover, integrating text-based retrieval with LLMs leverages their advanced language understanding capabilities to extract meaningful embeddings from textual data. For example, if a user frequently engages with science fiction movies, the system can identify themes such as ``futuristic technology'' or ``space exploration'' in item descriptions, facilitating recommendations of thematically similar films. Utilizing LLM embeddings in retrieval helps bridge the \textit{semantic gap} between user preferences and item descriptions.

Despite the advantages of text-based retrieval-augmented RS, their reliance on textual data introduces potential \textbf{security vulnerabilities}~\cite{deldjoo2022survey,chen2024black,chaudhari2024phantom,xue2024badrag,zou2024poisonedrag}. The open-ended nature of text-based metadata makes it susceptible to data poisoning attacks, where adversaries manipulate descriptions, tags, or other textual attributes to influence recommendations. For example, adversarial attacks can exploit these vulnerabilities to bias the exposure of certain items or manipulate the rankings of user classes~\cite{anelli2021msap}. Overall, the trustworthiness of LLM-based systems in recommendations have raised concerns about provider fairness, biases, and reliable and explainability, which must be considered when deploying these models~\cite{deldjoo2024understanding,deldjoo2025cfairllm}. In the context of RAG, data poisoning attacks could specifically target the retrieval process, exploiting the system's dependence on language-based representations to achieve undesirable outcomes, such as market manipulation or biased recommendations. For instance, tags—due to their lightweight and often user-generated nature—can be subtly altered in an adversarial manner, presenting opportunities for exploitation. The effectiveness of these manipulations depends on factors such as attack strategy, attacker knowledge (black-box vs. white-box), and system configuration~\cite{deldjoo2022survey}.


In this work, we investigate two types of adversarial poisoning attacks on retrieval-augmented RS (cf. Section~\ref{sec:proposed_method}):

\begin{itemize}
    \item \textbf{Local Adversarial Attacks}: These attacks personalize tag modifications for individual items by identifying similar items across different popularity classes using BERT embeddings across the entire dataset;
    \item \textbf{Global Adversarial Attacks}: These attacks apply a uniform pool of tags across the entire dataset;
\end{itemize}

\noindent Our analysis reveals that only the local tag selection method effectively manipulates recommendation outcomes, likely because its focused approach preserves semantic relevance and enhances adversarial impact. Overall, \textit{our work introduces a novel framework for attacking textual RAG systems and examining their vulnerabilities in provider-oriented evaluations (exposure-based and exposure-relevance), offering new insights and paving the way for future research in this area.} \\

\noindent \textbf{Contributions.} The contributions of this research are as follows:

\begin{itemize}
    \item We investigate the security vulnerabilities of RAG-based recommender systems by exploring adversarial data poisoning attacks that manipulate textual item metadata.
    \item We propose a novel methodology for selecting adversarial tags based on a combination of adversarial impact and semantic relevance, ensuring that the manipulated tags remain contextually appropriate and imperceptible to users.
    \item We conduct extensive experiments using the MovieLens dataset to evaluate the effectiveness of our attack strategies, analyzing their impact on recommendation outcomes and providing insights into the robustness of RAG-based systems.
\end{itemize}

\section{Poison-RAG: A Data Poisoning Attack Framework Against Retrieval Augmented Recommender Systems}

In this section, we introduce \texttt{Poison-RAG}, a data poisoning attack framework designed to manipulate item popularity in Retrieval-Augmented Generation (RAG)-based recommender systems. This framework focuses on designing attacks targeting the \textbf{producer side}, specifically by modifying item metadata—such as tags—to promote long-tail (unpopular) items and demote popular ones in recommendations.

We define the notation and variables used throughout this section in Table~\ref{tab:variables}.

\begin{table}[!b]
\centering
\caption{Summary of Notation}
\label{tab:variables}
\begin{tabular}{ll}
\toprule
Symbol & Description \\
\midrule
\(\mathcal{I}\) & Set of all items \\
\(\mathcal{U}\) & Set of all users \\
\(M_i\) & Metadata of item \(i\), including tags \\
\(R(u)\) & Set of items recommended to user \(u\) \\
\(R'(u)\) & Recommendations to user \(u\) after attack \\
\(\text{pop}(i)\) & Popularity class of item \(i\) (\textit{popular}, \textit{mid-tail}, \textit{long-tail}) \\
\(\mathcal{I}_L\) & Set of long-tail items \\
\(\mathcal{I}_P\) & Set of popular items \\
\(\Gamma\) & Set of adversarial tag modifications \\
\(\mathbb{I}(\cdot)\) & Indicator function (1 if condition is true, 0 otherwise) \\
\(A(t)\) & Adversarial impact of tag \(t\) \\
\(s(t, i)\) & Semantic relevance between tag \(t\) and item \(i\) \\
\(\mathbf{e}_t\) & Embedding vector of tag \(t\) \\
\(\mathbf{e}_i\) & Embedding vector of item \(i\) \\
\(\epsilon\) & Small constant to prevent division by zero \\
\(|R(u)|\) & Number of items recommended to user \(u\) \\
\bottomrule
\end{tabular}
\end{table}
\subsection{Attack Objectives and Setting}
\label{sec:attack_objectives}

Our attack scenario is summarized in Table~\ref{tab:attack_scenario}, outlining several key factors related to attack taxonomy, including the attacker’s knowledge level (black- vas. white-box data poisoning), power, and primary objectives within a given setting. The item meta data includes three main features: title, genre, and tag. Within this framework, the attacker modifies only \textit{tags} while keeping the genre and title unchanged to maintain the stealthiness of the attacks.

The success of the attack is evaluated based on shifts in popularity classes and alterations in item rankings within recommendation lists.

\vspace{2mm}
\noindent \textbf{Formalization of the Attack.} The adversary aims to identify a set of adversarial tag modifications, denoted as \(\Gamma\), that simultaneously:
\begin{itemize}
    \item \textbf{Maximizes the exposure of long-tail items} (\(i \in \mathcal{I}_L\)): Increasing their visibility and recommendation frequency.
    \item \textbf{Minimizes the exposure of popular items} (\(j \in \mathcal{I}_P\)): Reducing their prominence in recommendation outputs.
\end{itemize}

To quantify this, we define the \textit{exposure} of an item \(i\) to a user \(u\) as:
\[
\text{Exposure}(i, u) =
\begin{cases}
\mathbb{I}(i \in R'(u)), & \text{Binary Exposure} \\
\frac{1}{\log_2(\text{rank}_u(i) + 1)}, & \text{Continuous Exposure}
\end{cases}
\]
where:
\begin{itemize}
    \item \(\mathbb{I}(\cdot)\) is the indicator function.
    \item \(R'(u)\) represents the set of items recommended to user \(u\) after the attack.
    \item \(\text{rank}_u(i)\) denotes the position of item \(i\) in user \(u\)'s recommendation list.
\end{itemize}

The objective function of the adversary is formulated as:
\[
\max_{\Gamma} \frac{1}{|\mathcal{U}|} \sum_{u \in \mathcal{U}} \left[ \sum_{i \in R'(u) \cap \mathcal{I}_L} \text{Exposure}(i, u) - \sum_{j \in R'(u) \cap \mathcal{I}_P} \text{Exposure}(j, u) \right]
\]
This function seeks to maximize the cumulative exposure of long-tail items while minimizing that of popular items across all users. \vspace{2mm}

\noindent \textbf{Impact of Exposure Values.} In our framework, a \textit{high exposure} for long-tail items, (i.e., \(\text{Exposure}(i, u)\) is high for \(i \in \mathcal{I}_L\))indicates that these items are frequently present and/or appear near the top of recommendation lists, which contributes positively to the attack success. Conversely, a \textit{low exposure} value for popular items, i.e., (\(\text{Exposure}(j, u)\) is low for \(j \in \mathcal{I}_P\) suggests that these popular items are either absent or positioned lower in the recommendation lists, thus contributing again positively to the success of the attack.

\begin{table}[!t]
\centering
\caption{Situating Our Attack Scenario w.r.t. Literature}
\label{tab:attack_scenario}
\rowcolors{2}{rowgray}{white} 
\begin{tabular}{@{}p{2.7cm} p{9.7cm}@{}}
\toprule
\rowcolor{headerblue} \textbf{Aspect} & \textbf{Description} \\
\midrule
\textbf{Knowledge} & The attacker operates under a \textit{black-box setting,} lacking access to the internal parameters or architecture of the recommendation model. The attacker can observe the output recommendations by the system. \\
\midrule
\textbf{Capability} & The attacker can modify only item tags within the metadata, without altering other metadata (such as title and genre) or user interaction data. This limitation ensures the attack remains stealthy and inconspicuous (e.g., changing a movie title is not possible). \\
\midrule
\textbf{Primary Goals} & 
\textbf{1. Promote Long-Tail Items}: Enhance the visibility and recommendation frequency of long-tail (unpopular) items, shifting their popularity class, or raking upwards. \\
& \textbf{2. Demote Popular Items}: Decrease the exposure and ranking of popular items in the recommendation lists, reducing their prominence. \\
\midrule
\textbf{Success Metrics} & 
 \textbf{Pure Exposure-Based}: Observing changes in an item exposure, specifically shifts in its popularity class (e.g., moving from a long-tail to a mid-tail category), or its ranking position (continuous). \\
    
&  \textbf{Relevance Based}: Assessing changes in an item relevance (hit or ranking) within recommendations consider ground-truth, using per-class accuracy metrics, including NDCG, HR, and MRR.
\\
\bottomrule
\end{tabular}
\end{table}
\subsection{Adversarial Tag Selection Optimization}
\label{sec:proposed_method}

The optimization process for selecting adversarial tags is formalized as follows:
\[
\tilde{T}_i = \arg\max_{\tilde{T}_i \subseteq \mathcal{C}_i, \ |\tilde{T}_i| = k} \sum_{t \in \tilde{T}_i} A'(t, i)
\]
where:
\begin{itemize}
    \item \(\mathcal{C}_i\) is the candidate tag set for item \(i\), determined based on the chosen attack strategy (local or global).
    \item \(k\) is the number of tags to be modified for each item.
    \item \(A'(t, i)\) is the adversarial score for adding tag \(t\) to item \(i\),
\end{itemize}
\vspace{-2mm}
\subsubsection{Adversarial Tag Selection Method.} To achieve the attacker's goals, we design a method to select adversarial tags for items. The method balances two key factors:

\begin{enumerate}
    \item \textbf{Adversarial Impact \(A(t)\)}: Measures how strongly a tag \(t\) is associated with the desired popularity shift.
    \item \textbf{Semantic Relevance \(s(t, i)\)}: Ensures the selected tag \(t\) is contextually appropriate for item \(i\), maintaining imperceptibility.
\end{enumerate}

The overall adversarial score for adding tag \(t\) to item \(i\) is:

\[
A'(t, i) = A(t) \cdot s(t, i)
\]

Our goal is to select tags that maximize \(A'(t, i)\), thereby manipulating item popularity while keeping the modifications inconspicuous. \\

The adversarial impact of tag \(t\) is defined as:

\[
A(t) = \log \left( \frac{P(t \mid c_{\text{target}})}{P(t \mid c_{\text{orig}}) + \epsilon} \right)
\]

where \(P(t \mid c_{\text{target}})\) is the probability of tag \(t\) occurring in items of the adversarial target class (\textit{opposite} popularity class), and \(P(t \mid c_{\text{orig}})\) is the probability of tag \(t\) occurring in items of the original class. \(\epsilon\) is a small constant to prevent division by zero.

The probabilities are estimated using tag frequency counts:

\[
P(t \mid c) = \frac{\text{Count}(t, c) + \epsilon}{\sum_{t'} \text{Count}(t', c) + \epsilon \cdot |\mathcal{T}|}
\]

where \(\text{Count}(t, c)\) is the frequency of tag \(t\) in items of class \(c\), and \(|\mathcal{T}|\) is the total number of unique tags. Semantic relevance between tag \(t\) and item \(i\) is computed using cosine similarity between their embeddings:

\[
s(t, i) = \frac{\mathbf{e}_t^\top \mathbf{e}_i}{\|\mathbf{e}_t\| \|\mathbf{e}_i\|}
\]

\begin{itemize}
    \item \(\mathbf{e}_t\): Embedding vector of tag \(t\).
    \item \(\mathbf{e}_i\): Embedding vector of item \(i\), derived from its textual metadata.
\end{itemize}

\noindent Embeddings are obtained using pre-trained language models (BERT).

\subsection{Tag Selection Strategies.} To achieve differential exposure between long-tail and popular items, the attacker employs two distinct tag modification strategies: \textbf{local} and \textbf{global} tag selection. Both  trategies use a pool of tags as input, where the key difference lies in how these pools are constructed. The primary goal of each strategy is to \textit{select tags that can (subtly) shift an item popularity class or ranking while remaining undetectable, achieved by choosing tags from items that are semantically similar.}

In the \textit{local tag selection strategy}, a personalized pool of tags is constructed for each item by identifying similar items across different popularity classes using BERT embeddings. Tags are selected from this item-specific pool based on their adversarial impact and semantic relevance, ensuring that modifications remain contextually appropriate and inconspicuous. Conversely, the \textit{global tag selection strategy} employs a uniform pool of tags derived from items in the opposite popularity class across the entire dataset. While both strategies aim to manipulate recommendations, the local strategy focuses on item-specific similarities makes it more effective in achieving the desired popularity shift without raising suspicion.

\section{Experimental Setup}
\label{sec:setup}
\noindent \textbf{Data.} We use the MovieLens (\textit{``ml-latest-small''}) dataset, consisting of ratings, item information, and user-generated tags. To ensure data consistency, we retain only users with 20–100 interactions, excluding inactive users and superusers.

\begin{table}[h]
\caption{Statistics of the final dataset used in the experiment.}
\centering
\begin{tabular}{l S[table-format=4.0] S[table-format=4.0] S[table-format=6.0] S[table-format=5.2] S[table-format=4.2] S[table-format=2.2]}
\toprule
\textbf{Dataset} & \textbf{|U|} & \textbf{|I|} & \textbf{|R|} & \textbf{|R|/|U|} & \textbf{|R|/|I|} & \textbf{Density (\%)} \\
\midrule
\texttt{ml-latest} & 365 & 3079 & 16823 & 46.09 & 5.46 & 1.50 \\
\bottomrule
\end{tabular}
\label{table:dataset_stats}
\end{table}

\subsubsection{Data Enrichment and Augmentation Procedure.} We enrich the dataset by generating detailed item descriptions using GPT-3.5-turbo, enabling adversarial attack studies in both cold-start and non-cold-start scenarios. For each item, we prompt the LLM to create an engaging description based on its title and genres.

\begin{center}
\begin{tcolorbox}[colback=yellow!10, colframe=gray!80!black, title=Prompt for Description Augmentation]
\textbf{Prompt:} \\
Write a detailed and engaging description for the movie titled \texttt{\{title\}}, which falls under the genres \texttt{\{genres\}}. Use your knowledge about the movie to include important themes, characters, and plot points. Make the description appealing to potential viewers.
\end{tcolorbox}
\end{center}

\subsubsection*{Auto-Tagging NLP Steps.} We automatically generate relevant tags for each item based on the enriched descriptions using NLP techniques. This involves preprocessing the text, extracting significant keywords, and selecting the most relevant ones as tags.

\begin{table}[!h]
\caption{Example of data enrichment for a single item, showing the generated tags and enriched description.}
\label{table:enrichment_example}
\centering
\begin{tabular}{@{}llp{8cm}@{}}
\toprule
\textbf{Aspect} & \textbf{Before Enrichment} & \textbf{After Enrichment} \\
\midrule
\textbf{Title} & Planet Terror (2007) & Planet Terror (2007) \\
\textbf{Genres} & Action | Horror | Sci-Fi & Action | Horror | Sci-Fi \\
\textbf{Description} & Not available & \textit{`Planet Terror' is a gripping and blood-drenched tribute to classic grindhouse cinema, directed by Robert Rodriguez as a part of the double-feature `Grindhouse' with Quentin Tarantino's `Death Proof.' The film follows Cherry Darling, a go-go dancer turned fierce warrior, as she battles flesh-eating zombies and mutant creatures in a chaotic, post-apocalyptic world.} \\
\textbf{Tags} & None & \{cherry, zombie, survival, chaos, mutant\} \\
\bottomrule
\end{tabular}
\end{table}

\subsection{User Profile Building Strategy}

User profiles are constructed by aggregating item embeddings from each user's interaction history. Each item is represented by a GPT-3.5-turbo embedding vector that encapsulates its semantic attributes. We employ two aggregation methods: \textit{rating-weighted averaging} and \textit{temporal decay-weighted averaging}.

The rating-weighted averaging computes the user's embedding by taking a weighted average of the item embeddings, where weights are the user's ratings. The temporal decay-weighted averaging incorporates the recency of interactions by applying an exponential decay function \( w(t) = e^{-\lambda \cdot \text{time\_diff}(t)} \) with \( \lambda = 0.01 \), and a power factor \( \alpha = 1.2 \) is applied: \( w'(t) = w(t)^\alpha \).

By integrating these strategies, we capture both the intensity of user preferences and the temporal dynamics, resulting in robust user embeddings that enhance the recommendation system.
\section{Experiments and Results}
We evaluate the impact of adversarial attacks on both exposure and relevance metrics for a RAG-based recommender system. We compare recommendation outcomes before and after reranking and assess whether data augmentation (auto-tagging and generated content) mitigates attacks. We focus on two main research questions:

\begin{itemize}
    \item \textbf{RQ1:} What is the impact of adversarial attacks on item exposure at the retrieval stage?
    \item \textbf{RQ2:} Do adversarial attacks affect relevance metrics consistently before and after reranking, and does data generation help defend against these attacks?
\end{itemize}



\subsection*{RQ1: The Impact of Adversarial Attacks on Recommendation Exposure}
Table~\ref{tab:comparisons} reports exposure metrics (popularity lift, long-tail coverage, and mean popularity rank) under different scenarios. Notable cells in \textbf{red} cell indicate successful attacks (i.e., a stronger demotion of popular items or improved coverage for long-tail items), while cells in \textbf{blue} indicate failures (i.e., an unintended boost in popularity).

\begin{itemize}
    \item \textbf{Local vs. Global Attacks:} Local modifications effectively reduce popularity lift. For example, under the Decay strategy in cold-start conditions (Original i.e., no data augmentation), the baseline (\textbf{gray}) 0.9380 drops to 0.9191 (size~1, \textbf{red}) and 0.9282 (size~3, \textbf{red}), indicating \(\sim2\%\) and \(\sim1\%\) reductions. Global modifications, however, can accidentally boost popular items. For instance, popularity lift can rise to 1.3938 or 1.6235 (\textbf{blue}).
    
    \item \textbf{Data Augmentation:} With data augmentation (right side), local attacks still reduce popularity lift from a \textbf{gray} baseline of 1.3600 to 0.9304 and 0.9516 (both \textbf{red}). Global changes in augmented scenarios again raise popularity lift, confirming that global modifications are less effective for demoting popular items.
    
    \item \textbf{Long-Tail Coverage:} Most attack settings fail to substantially increase long-tail coverage. Although popular items are demoted, it remains difficult to meaningfully promote long-tail items.
\end{itemize}

\begin{tcolorbox}[colback=gray!10, colframe=white, title=Answer to RQ1]
Attacks can successfully demote popular items (especially through local tag modifications), but promoting long-tail items is much harder. Global modifications often backfire, inadvertently boosting popular items. Consequently, mid-tail items may end up promoted despite not being explicit attack targets.
\end{tcolorbox}

\begin{table}[!t]
\centering
\caption{Comparison of Popularity Lift, Long Tail Coverage, and Mean Popularity Rank in the Retrieval Stage. Values 1, 3, and 5 represent the size of altered tags in each attack scenario. The right panel shows results under data augmentation, where no item is in a complete cold-start scenario (cf. Section \ref{sec:setup}.) }
\label{tab:comparisons}
\hspace{-3cm} 
\footnotesize
\begin{adjustbox}{width=0.84\linewidth}
\begin{tabularx}{\textwidth}{ll|c|ccc|ccc|c|ccc|ccc}
\toprule
\multicolumn{16}{c}{\textbf{Retrieval stage}} \\
\midrule
\multirow{2}{*}{\textbf{Metric}} & \multirow{2}{*}{\textbf{Stage}} & \multicolumn{7}{c|}{\textbf{Attacked (Original)}} & \multicolumn{7}{c}{\textbf{Attacked (Data Augmentation)}} \\
\cmidrule(lr){3-9} \cmidrule(lr){10-16}
& & \multicolumn{1}{c}{B} & \multicolumn{3}{c}{Local} & \multicolumn{3}{c|}{Global} & \multicolumn{1}{c}{B} & \multicolumn{3}{c}{Local} & \multicolumn{3}{c}{Global} \\
\cmidrule(lr){4-6} \cmidrule(lr){7-9} \cmidrule(lr){11-13} \cmidrule(lr){14-16}
& & & \multicolumn{1}{c}{1} & \multicolumn{1}{c}{3} & \multicolumn{1}{c}{5} & \multicolumn{1}{c}{1} & \multicolumn{1}{c}{3} & \multicolumn{1}{c|}{5} & & \multicolumn{1}{c}{1} & \multicolumn{1}{c}{3} & \multicolumn{1}{c}{5} & \multicolumn{1}{c}{1} & \multicolumn{1}{c}{3} & \multicolumn{1}{c}{5} \\
\midrule

\multirow{2}{*}{\textbf{PLift$\downarrow$}} 
& Decay & \cellcolor{lightgray}0.9380 & \cellcolor{lightred}0.9191 & \cellcolor{lightred}0.9282 & 0.9260 & \cellcolor{lightblue}1.3938 & \cellcolor{lightblue}1.6235 & 1.6072 & \cellcolor{lightgray}1.3600 & \cellcolor{lightred}0.9304 & \cellcolor{lightred}0.9516 & 0.9420 & 0.9456 & 1.1994& 1.1614 \\
& Rating & 0.9314 & 0.8794 & 0.8785 & 0.8754 & 1.5420 & 1.8537 & 1.8409 & 1.4446 & 0.8907 & 0.9341 & 0.8857 &0.8981 & 1.1990 &1.1607 \\
\addlinespace

\multirow{2}{*}{\textbf{LT.Cov $\uparrow$}} 
& Decay & 0.0073 & 0.0046 & 0.0045 & 0.0045 & 0.0015 & 0.0012 & 0.0014 & 0.0040 & 0.0010 & 0.0010 & 0.0011 & 0.0010 & 0.0009 & 0.0012 \\
& Rating & 0.0086 & 0.0061 & 0.0061 & 0.0061 & 0.0003 & 0.0002 & 0.0003 & 0.0031 & 0.0004 & 0.0002 & 0.0002 & 0.0002 &0.0003 & 0.0003 \\
\addlinespace

\multirow{2}{*}{\textbf{P.Rank $\downarrow$}} 
& Decay & 26.961 & 26.418 & 26.679 & 26.616 & 40.062 & 46.6654 & 46.196 & 39.092 & 26.744 & 27.354 & 27.077  &27.182  & 34.478 & 33.385 \\
& Rating & 26.772 & 25.276 & 25.252 & 25.162 & 44.322 & 53.2808 & 52.9162 & 41.522 & 25.6034 & 26.8514 & 25.458& 25.815& 34.464 & 33.3652   \\
\addlinespace
\bottomrule
\end{tabularx}
\end{adjustbox}
\end{table}

\subsection*{RQ2: Consistency of Attacks Before and After Reranking, and the Effect of Data Generation}
Table~\ref{tab:consolidated_comparisons} compares relevance metrics (MRR, nDCG, and HR) both \textbf{before} and \textbf{after} reranking. Again, cells in \textbf{red} highlight a pronounced negative impact from the attack, while \textbf{blue} marks unexpected improvements favoring popular items.

\begin{itemize}
    \item \textbf{Consistent Impact on Popular Items:} In a generated (data-augmented) scenario, popular items' HR@1 decreases from a baseline of 0.3229 (\textbf{gray}) to 0.3021 (before reranking) and from 0.2708 (\textbf{gray}) to 0.2552 (after reranking). This represents a consistent drop of about 5--6\%. This shows a decrease in relevance, aligning with the attacker’s goal to \textit{demote popular items}.

    \item \textbf{Global Strategy Improves Popular Items' Relevance:} Only the global strategy shows higher relevance for popular items. HR@1 can rise from 0.3229 to 0.4010 or 0.4375 (\textbf{blue}), indicating a 24--35\% increase. This increase reflects an attack failure, as it inadvertently \textit{boosts the relevance of popular items}, contrary to the attacker’s intention. 
    
    \item \textbf{Long-Tail Items Remain Hard to Promote:} Long-tail item relevance generally declines (e.g., HR@1 drops from 0.0909 to 0.0303). Since the attack's goal is to promote long-tail items, this drop reflects an attack failure or more precisely \textit{ineffectiveness}.
    
    
    \item \textbf{Partial Defense via Data Augmentation:} Mid-tail items show stability or improvement in relevance metrics, especially under data augmentation (e.g., HR@1 rises from 0.3483 to 0.3652). These improvements are incidental, as mid-tail items were not explicitly targeted by the attacker. This does not directly represent attack success or failure, but does reflect partial defense due to augmentation.
\end{itemize}

\begin{tcolorbox}[colback=gray!10, colframe=white, title=Answer to RQ2]
Attacks consistently affect recommendation performance across retrieval and reranking. Global strategies can actually boost popular items' relevance, making the attack less detectible. While auto-tagging and data generation can slightly mitigate these issues (particularly for mid-tail items), they are not a complete defense against adversarial manipulation.
\end{tcolorbox}

\begin{table}[!h]
\centering
\caption{Relevance-based metric (Retrieval vs Augmentation)}
\label{tab:consolidated_comparisons}
\footnotesize
\hspace{-4cm}
\begin{threeparttable}
\begin{adjustbox}{width=0.76\textwidth}
\begin{tabularx}{\textwidth}{llc|c|ccc|ccc|c|ccc|ccc}
\toprule
\multicolumn{16}{c}{\textbf{Before Reranking}} \\
\midrule
\multirow{3}{*}{\textbf{Metric}} & \multirow{3}{*}{\textbf{Stage}} & \multirow{3}{*}{\textbf{Category}} & \multicolumn{7}{c|}{\textbf{Attacked}} & \multicolumn{7}{c}{\textbf{Attacked (Data Augmentation)}} \\
\cmidrule(lr){4-10} \cmidrule(lr){11-17}
& & & \multicolumn{1}{c}{B} & \multicolumn{3}{c}{Local} & \multicolumn{3}{c|}{Global} & \multicolumn{1}{c}{B} & \multicolumn{3}{c}{Local} & \multicolumn{3}{c}{Global} \\
\cmidrule(lr){5-7} \cmidrule(lr){8-10} \cmidrule(lr){12-14} \cmidrule(lr){15-17}
& & & & 1 & 3 & 5 & 1 & 3 & 5 & & 1 & 3 & 5 & 1 & 3 & 5 \\
\midrule

\multirow{3}{*}{\textbf{HR@k}} 
& Decay & mid tail & \cellcolor{lightgray}0.2753 & 0.3202 & 0.3202 & 0.3146 & 0.2472 & 0.1910 & 0.2135 & \cellcolor{lightgray}0.3483 & \cellcolor{white}0.3652 & 0.3652 & 0.3371 & 0.3539 & 0.3371 &0.3371 \\
& Decay & long tail & \cellcolor{lightgray}0.0909 & 0.0455 & 0.0455 & 0.0455 & \cellcolor{lightblue}0.0303 & \cellcolor{white}0.0606 & 0.0455 & \cellcolor{lightgray}0.0909 & 0.0455 & 0.0303 & 0.0455& 0.0455 & 0.0455 & 0.0455 \\
& Decay & popular & \cellcolor{lightgray}0.3229 & \cellcolor{lightred}0.3021 & 0.2864 & \cellcolor{lightred}0.2812 & \cellcolor{lightblue}0.4010 & \cellcolor{lightblue}0.4375 & 0.4375 & \cellcolor{lightgray}0.3750 & 0.1979 & 0.2083 & 0.3073 & 0.2031 & 0.3072 & 0.2917 \\
\addlinespace

\multirow{3}{*}{\textbf{MRR@k}} 
& Decay & mid tail & \cellcolor{lightgray}0.0522 & 0.0618 & \cellcolor{white}0.0621 & 0.0617 & 0.0459 & \cellcolor{white}0.0394 & 0.0414 & \cellcolor{lightgray}0.0746 & 0.0985 & \cellcolor{white}0.0984 &0.0951 &0.1029 &0.1023 & \cellcolor{white}0.0999 \\
& Decay & long tail & \cellcolor{lightgray}0.0080 & 0.0046 & 0.0045 & 0.0040 & 0.0034 & \cellcolor{white}0.0061 & 0.0044 & \cellcolor{lightgray}0.0081 & 0.0039 & 0.0031 & 0.0039 & 0.0039 & 0.0035 & \cellcolor{white}0.0035 \\
& Decay & popular & \cellcolor{lightgray}0.0816 & 0.0696 & \cellcolor{white}0.0687 & 0.0684 & 0.1154 & \cellcolor{white}0.1099 & 0.1097 & \cellcolor{lightgray}0.0816 & 0.0385 & 0.0486 & 0.0409 & 0.0408 & 0.0671 & \cellcolor{white}0.0664 \\
\addlinespace

\multirow{3}{*}{\textbf{nDCG@k}} 
& Decay & mid tail & \cellcolor{lightgray}0.0311 & 0.0365 & 0.0365 & 0.0358 & 0.0273 & 0.0232 & 0.0241 & \cellcolor{lightgray}0.0308 & 0.0402 & 0.0400 & 0.0379& 0.0394 & 0.0365 &0.0365\\
& Decay & long tail & \cellcolor{lightgray}0.0179 & 0.0092 & 0.0091 & 0.0088& 0.0078 & 0.0118 & 0.0096 & \cellcolor{lightgray}0.0151 & 0.0094 & 0.0072 & 0.0093 & 0.0093 & 0.0090 & 0.0091 \\
& Decay & popular & \cellcolor{lightgray}0.0287 & 0.0251 & 0.0241 & 0.0234 & 0.0493 & 0.0546 & 0.0533 & \cellcolor{lightgray}0.0385 & 0.0157 & 0.0196 & 0.0192 & 0.0167 & 0.0277 & 0.0275 \\
\midrule

\multicolumn{16}{c}{\textbf{After Re-ranking (Cutoff = 10)}} \\
\midrule
\multirow{3}{*}{\textbf{Metric}} & \multirow{3}{*}{\textbf{Stage}} & \multirow{3}{*}{\textbf{Category}} & \multicolumn{7}{c|}{\textbf{Attacked}} & \multicolumn{7}{c}{\textbf{Attacked (Data Augmentation)}} \\
\cmidrule(lr){4-10} \cmidrule(lr){11-17}
& & & \multicolumn{1}{c}{B} & \multicolumn{3}{c}{Local} & \multicolumn{3}{c|}{Global} & \multicolumn{1}{c}{B} & \multicolumn{3}{c}{Local} & \multicolumn{3}{c}{Global} \\
\cmidrule(lr){5-7} \cmidrule(lr){8-10} \cmidrule(lr){12-14} \cmidrule(lr){15-17}
& & & & 1 & 3 & 5 & 1 & 3 & 5 & & 1 & 3 & 5 & 1 & 3 & 5 \\
\midrule

\multirow{3}{*}{\textbf{HR@k}} 
& Decay & mid tail & \cellcolor{lightgray}0.2360 & 0.2584 & 0.2528 & 0.2528 & 0.1966 & 0.1629 & 0.1685 & \cellcolor{lightgray}0.2697 & 0.3258 & 0.3146 &0.3089 & 0.3258 & 0.2921 &0.2921\\
& Decay & long tail & \cellcolor{lightgray}0.0758 & 0.0303 & 0.0303 & 0.0303 & 0.0303 & 0.0606 & 0.0455 & \cellcolor{lightgray}0.0909 & 0.0303 & 0.0151 & 0.0303 & 0.0303 & 0.0303 & 0.0303 \\
& Decay & popular & \cellcolor{lightgray}0.2708 & \cellcolor{white}0.2552 & 0.2500 & 0.2395 & 0.3593 & 0.4062 & 0.3802 & \cellcolor{lightgray}0.3594 & 0.1719 & 0.1875 & 0.1979 & 0.1822 & 0.2864& 0.2604 \\
\addlinespace

\multirow{3}{*}{\textbf{MRR@k}} 
& Decay & mid tail & \cellcolor{lightgray}0.1246 & 0.1411 & 0.1472 & 0.1519 & 0.1101 & 0.0934 & 0.0862 & \cellcolor{lightgray}0.1360 & 0.2081 & 0.2030 & 0.1964 & 0.2112 & 0.1851 &0.1942 \\
& Decay & long tail & \cellcolor{lightgray}0.0434 & 0.0152 & 0.0227 & 0.0227 & 0.0126 & 0.0454 & 0.0379 & \cellcolor{lightgray}0.0476 & 0.0101 & 0.0075 & 0.0101 & 0.0101 & 0.0126 & 0.0126 \\
& Decay & popular & \cellcolor{lightgray}0.1581 & 0.1438 & 0.1434 & 0.1339 & 0.2101 & 0.2329 & 0.2408 & \cellcolor{lightgray}0.1876 & 0.0910 & 0.0980 & 0.1218 & 0.1109 & 0.1456 & 0.1365 \\
\addlinespace

\multirow{3}{*}{\textbf{nDCG@k}} 
& Decay & mid tail & \cellcolor{lightgray}0.0613 & 0.0699 & 0.0698 & 0.0710 & 0.0528 & 0.0462 & 0.0463 & \cellcolor{lightgray}0.0567 & 0.0879 & 0.0872 & 0.0826 & 0.08511 & 0.0741 & 0.0805\\
& Decay & long tail & \cellcolor{lightgray}0.0298 & 0.0153 & 0.0197 & 0.0187& 0.0141 & 0.0299 & 0.0263 & \cellcolor{lightgray}0.0308 & 0.0124 &0.0094 & 0.0123 & 0.0123 & 0.0141 & 0.0141 \\
& Decay & popular & \cellcolor{lightgray}0.0533 & 0.0508 & 0.0504 & 0.0470 & 0.0861 & 0.0964 & 0.0979 & \cellcolor{lightgray}0.0789 & 0.0319 & 0.0368 & 0.0416 & 0.0374 & 0.0547 & 0.0531 \\
\bottomrule
\end{tabularx}
\end{adjustbox}
\end{threeparttable}
\end{table}

\subsection{Related Work}
Adversarial machine learning research spans poisoning and evasion attacks, targeting either model parameters or inputs~\cite{deldjoo2024review,fan2024survey}. Retrieval-Augmented Generation (RAG) systems face specific vulnerabilities when malicious text is injected into external knowledge sources~\cite{chen2024black,chaudhari2024phantom,xue2024badrag,zou2024poisonedrag}. Recent works like BadRAG~\cite{xue2024badrag} and PoisonedRAG~\cite{zou2024poisonedrag} show that an attacker can bias generative outputs by altering retrieved content. Another line of research focuses on privacy vulnerabilities in RAG systems, with techniques such as Membership Inference Attacks (MIAs) assessing whether specific data exists within the retrieval database~\cite{anderson2024my,li2024generating}.

Despite these efforts, most prior research targets general-purpose RAG systems rather than recommender-specific implementations. Our work extends this area by focusing on adversarial risks in RAG-based recommendation pipelines. We introduce \textit{Poison-RAG}, which manipulates textual metadata to influence recommendations by demoting popular items and assessing the challenges of promoting long-tail items. Unlike prior work, we also propose a data augmentation strategy that includes auto-tagging and enriched content generation, showcasing its potential to mitigate such attacks. These contributions provide a novel perspective on securing RAG-based recommendation systems.

\section{Conclusion and Future Works}
This study examines the vulnerability of \textbf{RAG-based recommender systems} to adversarial data poisoning through textual metadata manipulation. The proposed \textbf{Poison-RAG} framework demonstrates that while demoting popular items is relatively feasible, promoting long-tail items remains challenging due to their inherent resistance to manipulation. Experimental results highlight that local attack strategies are more effective than global ones, with the latter often inadvertently boosting popular items' relevance. Additionally, data augmentation strategies, such as auto-tagging and enriched metadata generation, offer partial resilience, particularly for mid-tail items.

Future work could explore hybrid attack-defense mechanisms, focusing on balancing robustness against adversarial manipulation with maintaining recommendation quality. Strengthening defenses against adversarial risks in RAG systems is crucial to ensure secure and trustworthy recommendations.
\bibliographystyle{splncs04}
\bibliography{refs}
\end{document}